\documentclass[aps, superscriptaddress, floatfix]{revtex4}
\usepackage[english]{babel}
\usepackage{epsfig}
\usepackage{changebar}
\usepackage{graphicx}

\begin{document}
\renewcommand{\figurename}{{\bf Fig.}}
\renewcommand{\tablename}{{\bf Tab.}}
\title{Quark-Antiquark Condensates in the Hadronic Phase}

\author{A.~Tawfik}
\affiliation{Faculty of Physics, University of Bielefeld, D-33501
  Bielefeld, Germany}
\author{D.~Toublan}
\affiliation{Physics Department, University of Illinois at Urbana-Champaign,
Urbana, IL 61801} 

\date{\today}

\begin{abstract}
We use a hadron resonance gas model to calculate the quark-antiquark
condensates for light (up and down) and strange quark flavors at finite
temperatures and chemical potentials. 
At zero chemical potentials, we find that at the temperature where
the light quark-antiquark condensates entirely vanish the strange
quark-antiquark condensate still keeps a relatively large fraction of its
value in the vacuum. This is in agreement with results obtained in
lattice simulations and in chiral perturbation theory at finite
temperature and zero chemical potentials. Furthermore, we find that this
effect slowly disappears at larger baryon chemical potential. 
These results might have significant
consequences for our understanding of QCD at finite
temperatures and chemical potentials. 
Concretely, our results imply that there might be  a domain of
temperatures where  chiral symmetry is restored for light quarks, but still
broken for strange quark that persists at small chemical potentials.
This might have practical consequences for heavy ion collision
experiments. 
\end{abstract}

\maketitle

\section{\label{sec:1}Introduction}

There are many important physical systems where the strong interaction at
nonzero temperature and density plays a crucial role: e.g.,  neutron
stars,  heavy-ion collision experiments, and the early universe. To
understand these systems, we need to better grasp the phase diagram 
of QCD at nonzero temperature and chemical potentials.  Unfortunately, the most
reliable techniques used at nonzero temperature and zero chemical
potentials are generally not applicable at finite chemical potentials.  In
particular, lattice simulations suffer from the so-called
sign problem.  This problem has not yet been solved
explicitly.  However, some recent advances have allowed the study of  the
high temperature and low chemical potential part of the phase
diagram~\cite{lattMuB_F&K, lattMuB_Bielefeld, lattMuB_ZH, lattMuB_Maria, 
  lattMuB_Gupta, lattMuB_Azcoiti}. Our current understanding of the phase
diagram is therefore mainly based on various models which validity is difficult
to test~\cite{cscRev}.  In this article we shall use a hadron resonance
gas model to study properties of the QCD phase diagram at nonzero
temperature and chemical potentials. It is one of the models that has
been successfully used in the past 
to study the hadronic phase at finite temperature and 
chemical potentials~\cite{hicRev, hgrExp, hgrLatt, hgrTc}. 
This model is in good quantitative agreement
with the lattice simulations for various observables in the hadronic
phase~\cite{hgrLatt}. Furthermore, it has also been used to determine the
critical temperature that separates the hadronic phase from the quark-gluon
plasma phase~\cite{hgrTc}. We would like to stress here again that the
agreement with the lattice simulations is excellent, especially at small
chemical potential~\cite{lattMuB_F&K, lattMuB_Bielefeld, lattMuB_ZH,
  lattMuB_Maria}. 

Recent lattice simulations at finite temperatures and zero chemical
potentials seem to indicate that the critical temperature for the light
quark-antiquark condensate is different than that for the strange
quark-antiquark condensate~\cite{MILCthermo}.  In particular, the
lattice results for the different quark number
susceptibilities indicate that the pseudocritical temperature of the
crossover for the restoration of chiral symmetry for the light 
quarks is smaller than that for the strange quark \cite{MILCthermo}.
This difference has 
also been observed in chiral perturbation theory at zero chemical
potential~\cite{pelaez}. 
In this article  we shall use the hadron resonance gas model to study the
possible differences between the behavior of the light
$<\bar{q}q>=<\bar{u}u>=<\bar{d}d>$ and strange $<\bar{s}s>$  
quark-antiquark condensates in the hadronic phase. 
At small chemical potentials, we find that the strange quark-antiquark
condensate is still large where the light quark-antiquark condensates become
very small.  At higher chemical potentials, this difference slowly
diminishes: The light and strange quark-antiquark condensates 
become small at the same temperature.  The difference between the
light and strange quark-antiquark condensates might lead to
significant consequences for the heavy-ion collision experiments. One can
expect the existence of a domain of temperatures where the chiral symmetry is
restored for the light quark flavors and broken for the strange quark flavor.

\section{\label{sec:2}Formalism}

The pressure in the hadronic phase
is given by the contributions of all the hadron resonances up to $2$~GeV
treated as a free gas~\cite{hgrLatt, hgrTc}. All thermodynamic quantities
can be derived from the pressure  
\begin{eqnarray}
  \label{pZ}
  p = \lim_{V\rightarrow\infty} \, \frac TV \, \ln\,Z(T,\mu_B,\mu_I,\mu_S, V),
\end{eqnarray}
where $ Z(T,\mu_B,\mu_I,\mu_S,V)$ is the grand canonical partition
function in a finite volume $V$, at nonzero temperature, $T$, baryon
chemical potential, $\mu_B$, isospin chemical potential, $\mu_I$, and
strangeness chemical potential, $\mu_S$. It has been found that this model
gives a very good description of the hadronic phase and of the critical
temperature that separates the hadronic phase from the quark-gluon plasma
phase~\cite{hgrLatt, hgrTc}. 

In the free gas approximation, the contribution to the pressure due to
a particle of mass $m_h$, baryon charge $B$, isospin $I_3$, strangeness $S$, 
and degeneracy $g$ is given by
\begin{eqnarray}
  \label{p1}
  \Delta p=\frac{g\, m_h^2\, T^2}{2\pi^2} \,
\sum_{n=1}^\infty \,\frac{(-\eta)^{n+1}}{n^2}
\,\exp\left(n\frac{B\mu_B - I_3\mu_I - S \mu_S}{T} \right)
\, K_2\left(n\frac{m_h}{T} \right),
\end{eqnarray}
where $\eta=+1$ for fermions and $\eta=-1$ for bosons. $K_n(x)$ is
the modified Bessel function. In the hadronic phase the isospin is
an almost exact symmetry since the hadronic spectrum is almost isospin
symmetric~\cite{hgrTc}.    

The quark-antiquark condensates are given by the
derivative of the pressure with respect to the constituent quark masses:
\begin{eqnarray}
\label{qqHRG}
<\bar{q}q>&=&<\bar{q}q>_0+ \sum_h \frac{\partial
  m_h}{\partial m_q} 
\frac{\partial \Delta p}{\partial m_h}, \nonumber \\ 
<\bar{s}s>&=&<\bar{s}s>_0+ \sum_h 
\frac{\partial m_h}{\partial m_s} \frac{\partial \Delta
  p}{\partial m_h}, 
\end{eqnarray}
where $<\bar{q}q>=<\bar{u}u>=<\bar{d}d>$ represents the light quark-antiquark
condensate. $<\bar{q}q>_0$ and $<\bar{s}s>_0$ indicate the value
of the light and strange quark-antiquark condensates in the vacuum,
respectively. 

The computation of the quark-antiquark condensates in Eq.~(\ref{qqHRG})
requires modeling two quantities. The first one is the  strange
quark-antiquark condensate at zero temperature and zero chemical potential,
$<\bar{s}s>_0$. For it we need to  
know the kaon decay constant, which is unfortunately not yet known
experimentally. But in the framework of lattice QCD, the
ratio of heavy to light meson decay constants is found to be given by
$F_K/F_{\pi}=1.16\pm0.04$~\cite{lattHeavy} 
(see also \cite{LeutwylerRoos}). Using QCD sum rules, the
ratio of strange to light quark-antiquark condensates in vacuum is given by
$0.8\pm0.3$~\cite{jamin} (another estimation gives
$0.75\pm0.12$~\cite{narison}).  The Gell-Mann--Oakes--Renner relation connects
the quark-antiquark condensates at zero temperature and zero chemical 
potential to the meson masses and to their decay constants. We use the
next-to-leading order result obtained in chiral perturbation theory
\cite{GLsu3}: 
\begin{eqnarray}
\label{GOR}
F_\pi^2 m_\pi^2 \left(1-\kappa \frac{m_\pi^2}{F_\pi^2}\right)
 &=& (m_u+m_d) <qq>_0, \nonumber \\
F_K^2 m_K^2 \left(1-\kappa \frac{m_K^2}{F_\pi^2}\right) &=& \frac12 (m_q + m_s)
(<\bar{q}q>_0 + <\bar{s}s>_0),   
\end{eqnarray}
where $m_q$ stands for the light quark mass where we assume
$m_u=m_d$. The coefficient 
$\kappa=0.021\pm0.008$ has been obtained from the low-energy
coupling constants of chiral perturbation
theory~\cite{GLsu3,jamin}. 
Notice that $\kappa$ does not contain any chiral logarithm.

The second quantity which we have to model in order to calculate the
quark-antiquark condensates in the hadron gas model Eq.~(\ref{qqHRG}) is
the quark mass dependence of the hadron masses. Results from lattice
simulations~\cite{hgrLatt} indicate that
\begin{eqnarray}
\label{lattMass}
\frac{\partial m_h}{\partial (m_\pi^2)}&=&\frac{A}{m_h},  
\end{eqnarray}
where $A\sim0.9 - 1.2$.  Using these results together with the
Gell-Mann--Oakes--Renner relation, Eq.~(\ref{GOR}), one finds
that~\cite{hgrTc}  
\begin{eqnarray}
\frac{\partial m_h}{\partial m_q} &\equiv& 
\frac{\partial (m_\pi^2)}{\partial  m_q}  \frac{\partial
  m_h}{\partial (m_\pi^2)}  = 
\frac{A\;<\bar{q}q>_0 \left(1+2 \kappa \frac{m_\pi^2}{F_\pi^2}\right)}
{F_{\pi}^2\;m_h}. \label{mhmq} \\   
\frac{\partial m_h}{\partial m_s} &\equiv& \frac{\partial (m_K^2)}{\partial
  m_s} \frac{\partial (m_\pi^2)}{\partial (m_K^2)} \frac{\partial
  m_h}{\partial (m_\pi^2)} =
\frac{A\;<\bar{q}q>_0 \left(1+2 \kappa \frac{m_\pi^2}{F_\pi^2}\right)}
{F_{\pi}^2\;m_h}. \label{mhms}  
\end{eqnarray}
Relation~(\ref{mhms}) is not valid for the pion mass, since pions are
almost independent of $m_s$ in chiral perturbation
theory~\cite{GLsu3}. 
However, relations (\ref{mhmq}) and (\ref{mhms})
 work reasonably well for the nucleon~\cite{Sainio}, for instance. 
We therefrom assume that this relation is 
valid for all the hadrons heavier than the pions. This is in agreement with
the lattice simulations, where the masses of a few hadrons have been shown to
follow Eq.~(\ref{mhmq}) and Eq.~(\ref{mhms}) over a sizable range of quark
masses \cite{hgrLatt}. 

Using these assumptions, we find that the contribution of one hadron of
mass $m_h$ to the light quark-antiquark condensates is given by
\begin{eqnarray}
  \label{qq}
  \frac{\Delta <\bar{q}q>}{<\bar{q}q>_0} &=& 
-\frac{g }{2\pi^2}\;T\, m_h\; \frac{A (1+2 \kappa
  \frac{m_\pi^2}{F_\pi^2})}{F_\pi^2} \nonumber \\
& & \sum_{n=1}^\infty \,\frac{(-\eta)^{n+1}}{n}
\,\exp\left(n\frac{B\mu_B -I_3\mu_I-S\mu_S}{T} \right)
\, K_1\left( n\frac{m_h}{T} \right),
\end{eqnarray}
whereas its contribution to the strange quark-antiquark condensate is
given by
\begin{eqnarray}
  \label{ss}
  \frac{\Delta <\bar{s}s>}{<\bar{s}s>_0} &=& 
-\frac{g}{2\pi^2} \;T\, m_h\;
\frac{A (1+2 \kappa
  \frac{m_\pi^2}{F_\pi^2})}{F_\pi^2}\,\frac{<\bar{q}q>_0}{<\bar{s}s>_0} 
\nonumber \\
& & \sum_{n=1}^\infty \,\frac{(-\eta)^{n+1}}{n}
\,\exp\left(n\frac{B\mu_B -I_3\mu_I-S\mu_S}{T} \right)
\, K_1\left(n\frac{m_h}{T} \right).
\end{eqnarray}

\section{\label{sec:3}Results}

\begin{figure}[h]
\includegraphics*[scale=0.55, clip=true, angle=0,
draft=false]{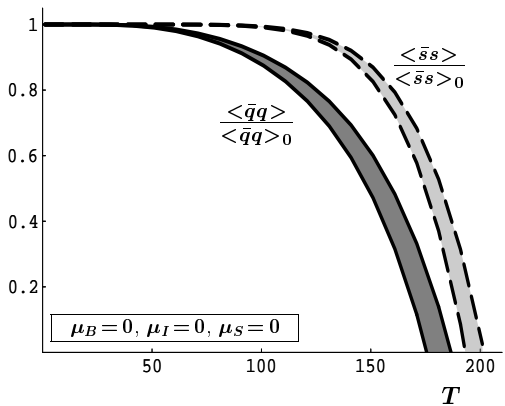}
\caption{\label{fig1} The quark-antiquark condensates as a function of
  temperature $T$ at zero baryon, isospin, and strangeness chemical
  potentials. The light quark-antiquark condensate
  $\frac{<\bar{q}q>}{<\bar{q}q>_0}$ is in dark
gray and solid curve. The strange quark-antiquark condensate
$\frac{<\bar{s}s>}{<\bar{s}s>_0}$ is given in light gray and
dashed curve.} 
\end{figure}

\begin{figure}[h]
\includegraphics*[scale=0.55, clip=true, angle=0,
draft=false]{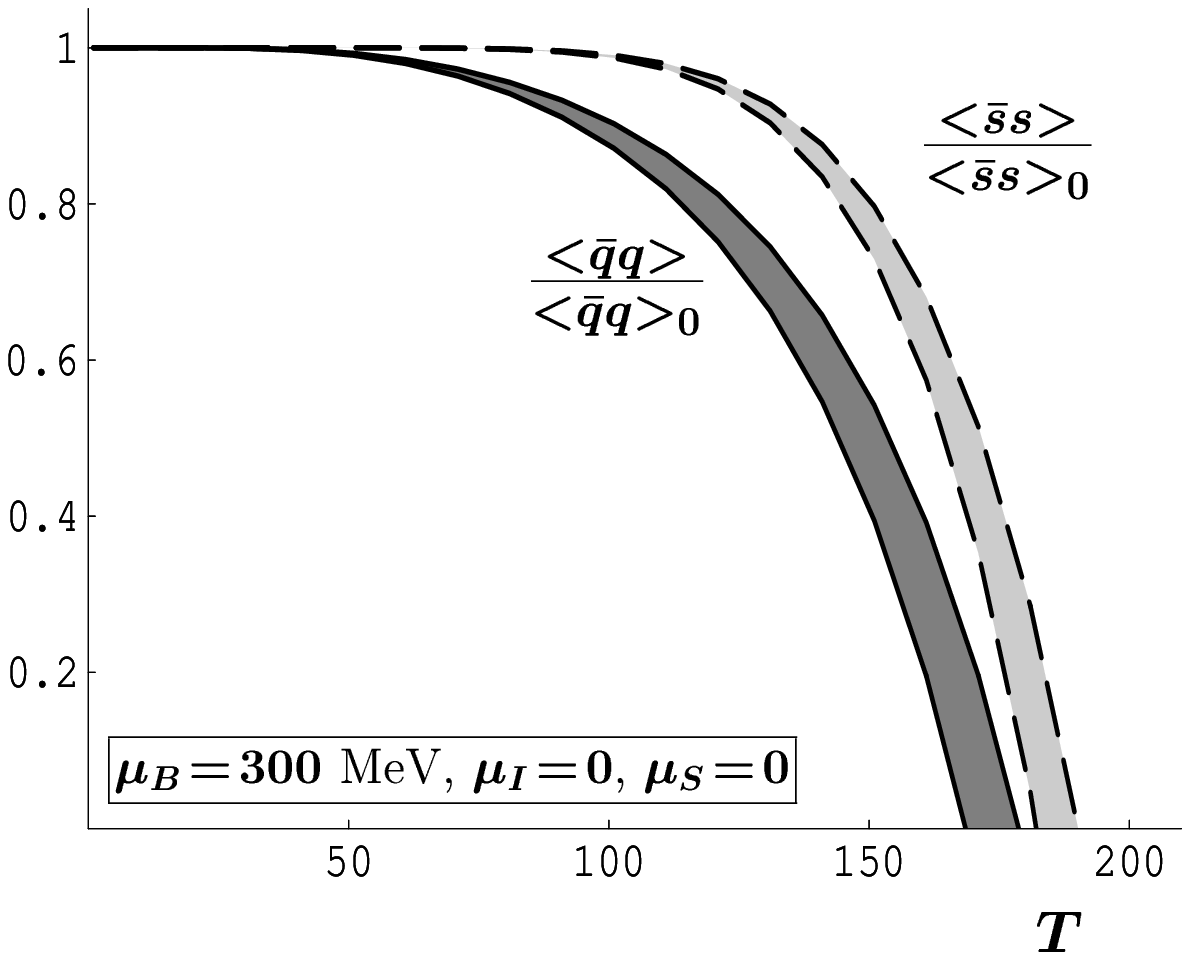}
\hspace{1.4cm}
\includegraphics*[scale=0.55, clip=true, angle=0,
draft=false]{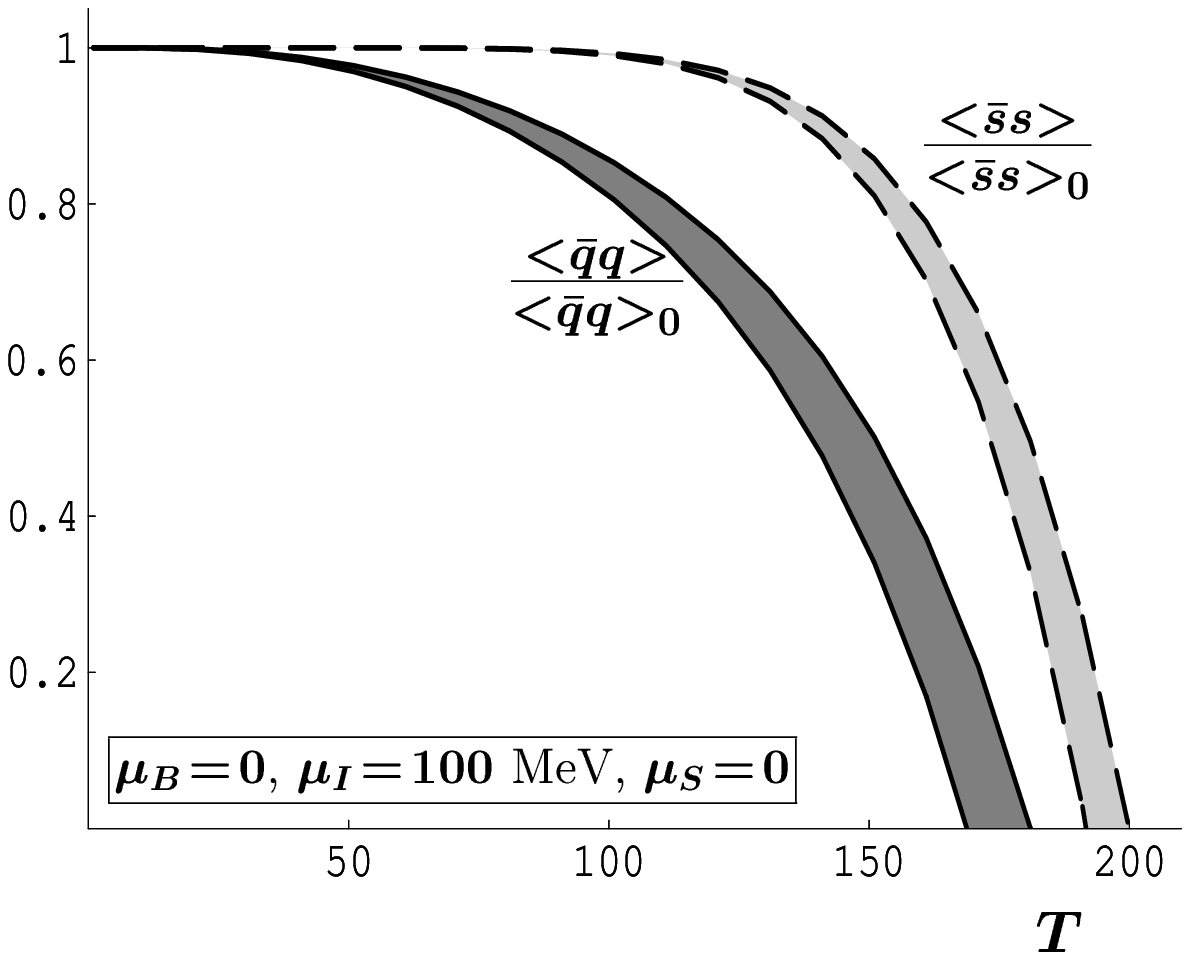}
\vspace{.3cm}
\includegraphics*[scale=0.55, clip=true, angle=0,
draft=false]{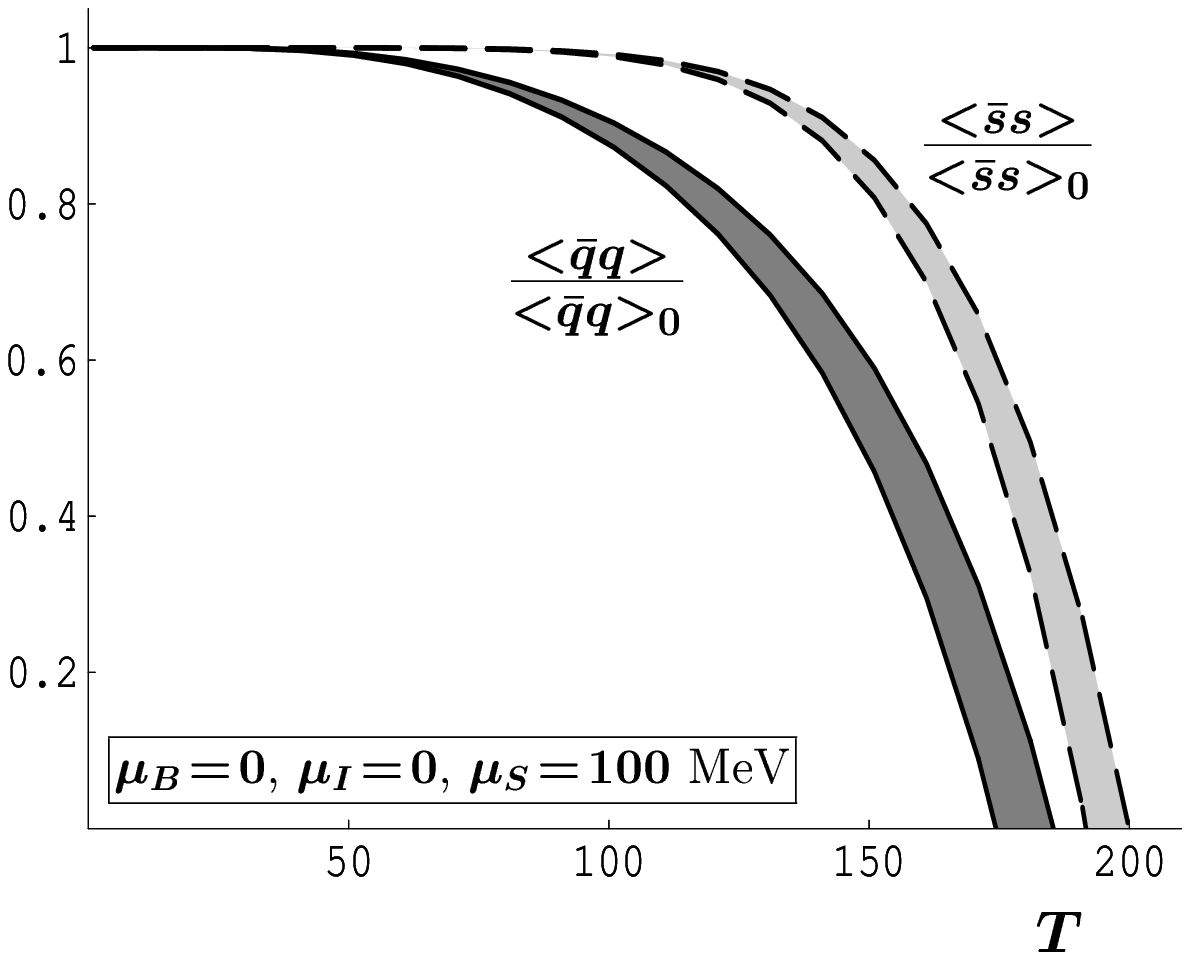}
\hspace{1.4cm}
\includegraphics*[scale=0.55, clip=true, angle=0,
draft=false]{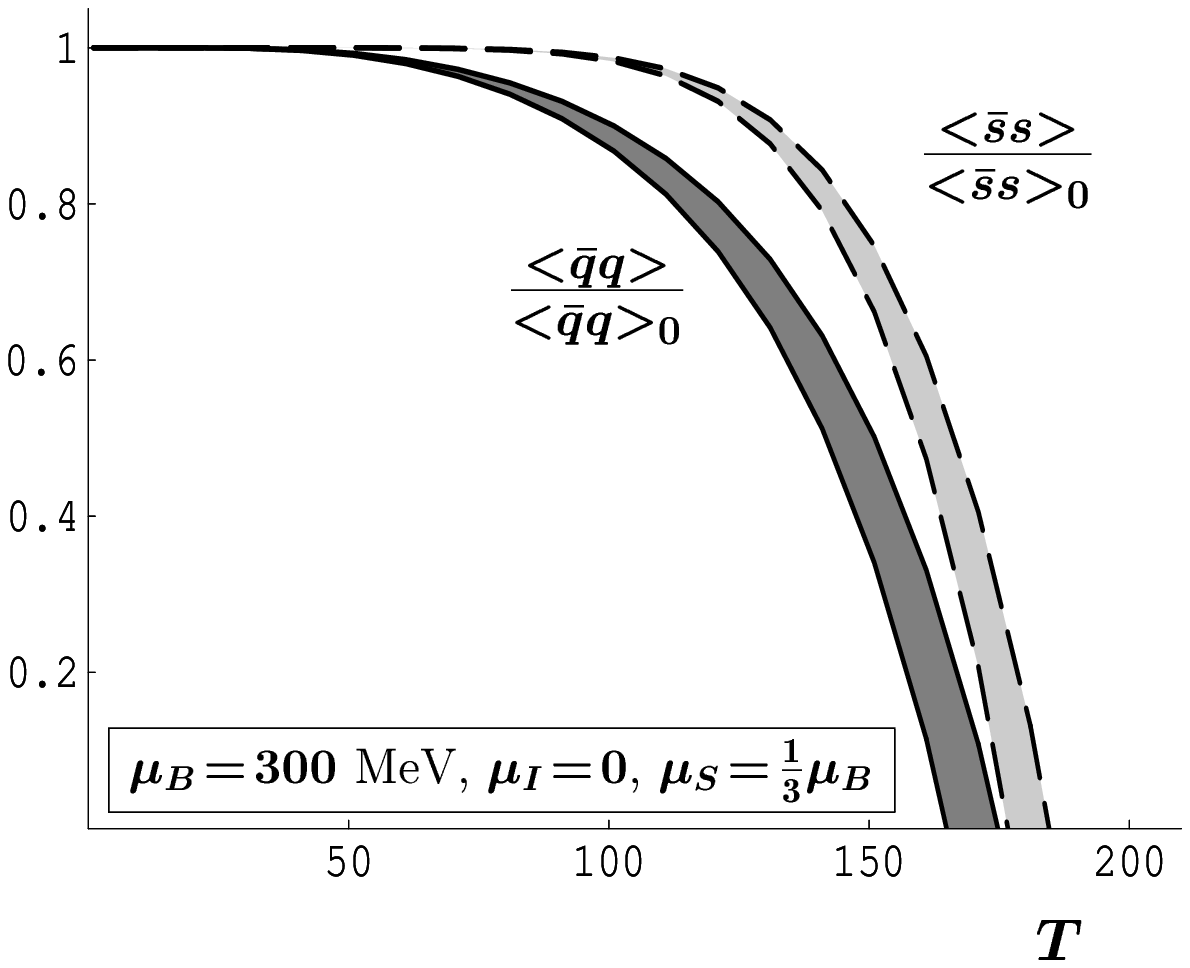}
\caption{\label{fig2} The quark-antiquark condensates as a function of
  temperature at different values of baryon, isospin, and strangeness
  chemical potentials.  
The light quark-antiquark condensate $\frac{<\bar{q}q>}{<\bar{q}q>_0}$
is in dark
gray and solid curve. The strange quark-antiquark condensate
$\frac{<\bar{s}s>}{<\bar{s}s>_0}$ is given in light gray and
dashed curve. 
Upper-left panel: $\mu_B=300$~MeV, $\mu_I=0$, and $\mu_S=0$.
Upper-right panel: $\mu_B=0$, $\mu_I=100$~MeV, and $\mu_S=0$.
Lower-left panel: $\mu_B=0$, $\mu_I=0$, and $\mu_S=100$~MeV.
Lower-right panel: $\mu_B=300$~MeV, $\mu_I=0$, and $\mu_S=100$~MeV.
Although chemical potentials have been set to finite values, the critical
  temperature of light quark condensate is still smaller than that of the
  strange quark one. }
\end{figure}

We include all hadron resonances with masses up to $\sim2$~GeV.  
The light and strange quark-antiquark condensates,
Eq.~(\ref{qq})~and~Eq.~(\ref{ss}) are calculated as a function of 
temperature  $T$ and for various values of 
baryon, isospin, and strangeness chemical potentials, $\mu_B,$,
$\mu_I$, and $\mu_D$, respectively.  The results are shown
in Fig.~\ref{fig1} and~Fig.~\ref{fig2}. 

\begin{figure}[h]
\includegraphics*[scale=0.45, clip=true, angle=0,
draft=false]{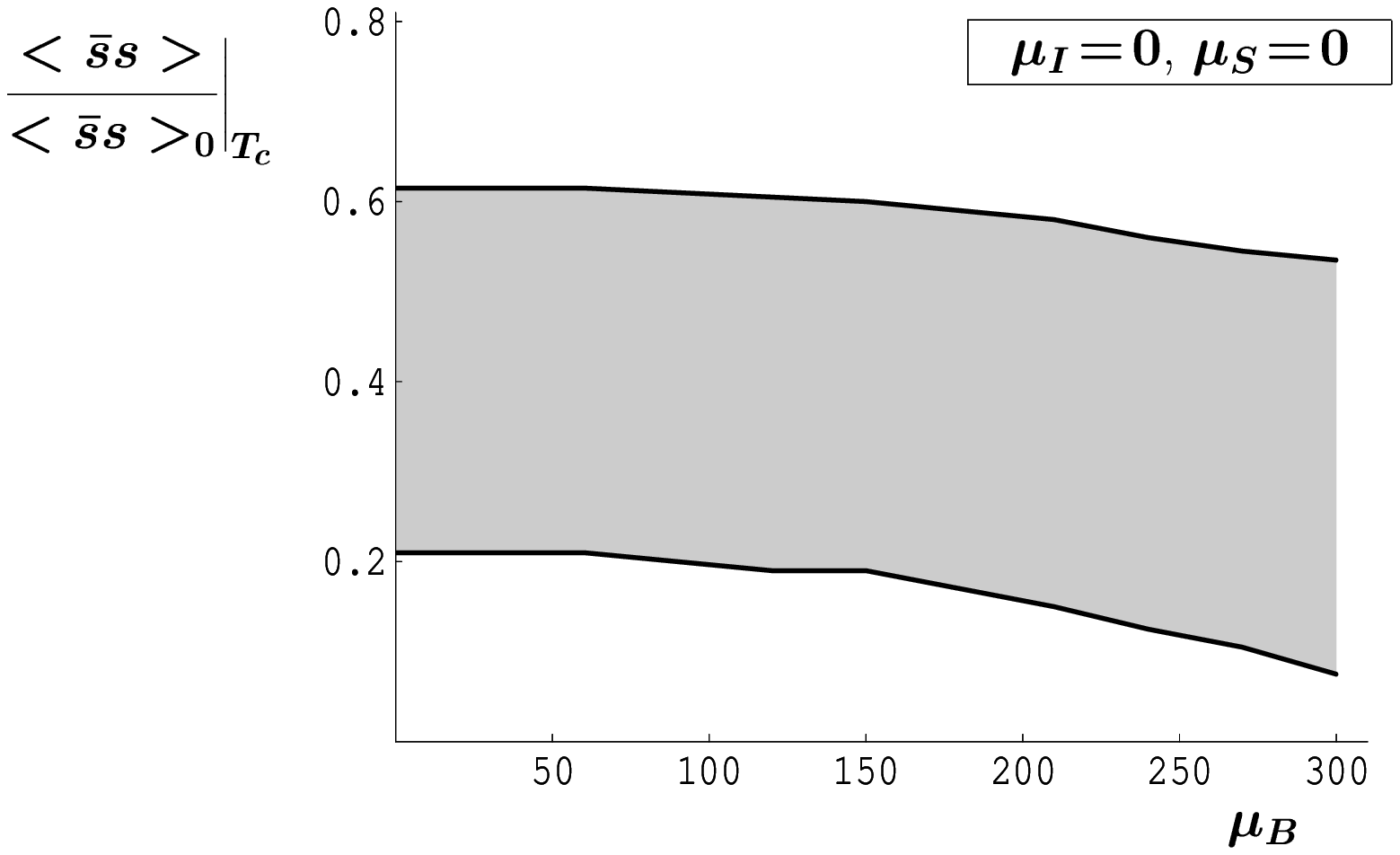}
\hspace{2cm}\includegraphics*[scale=0.45, clip=true, angle=0,
draft=false]{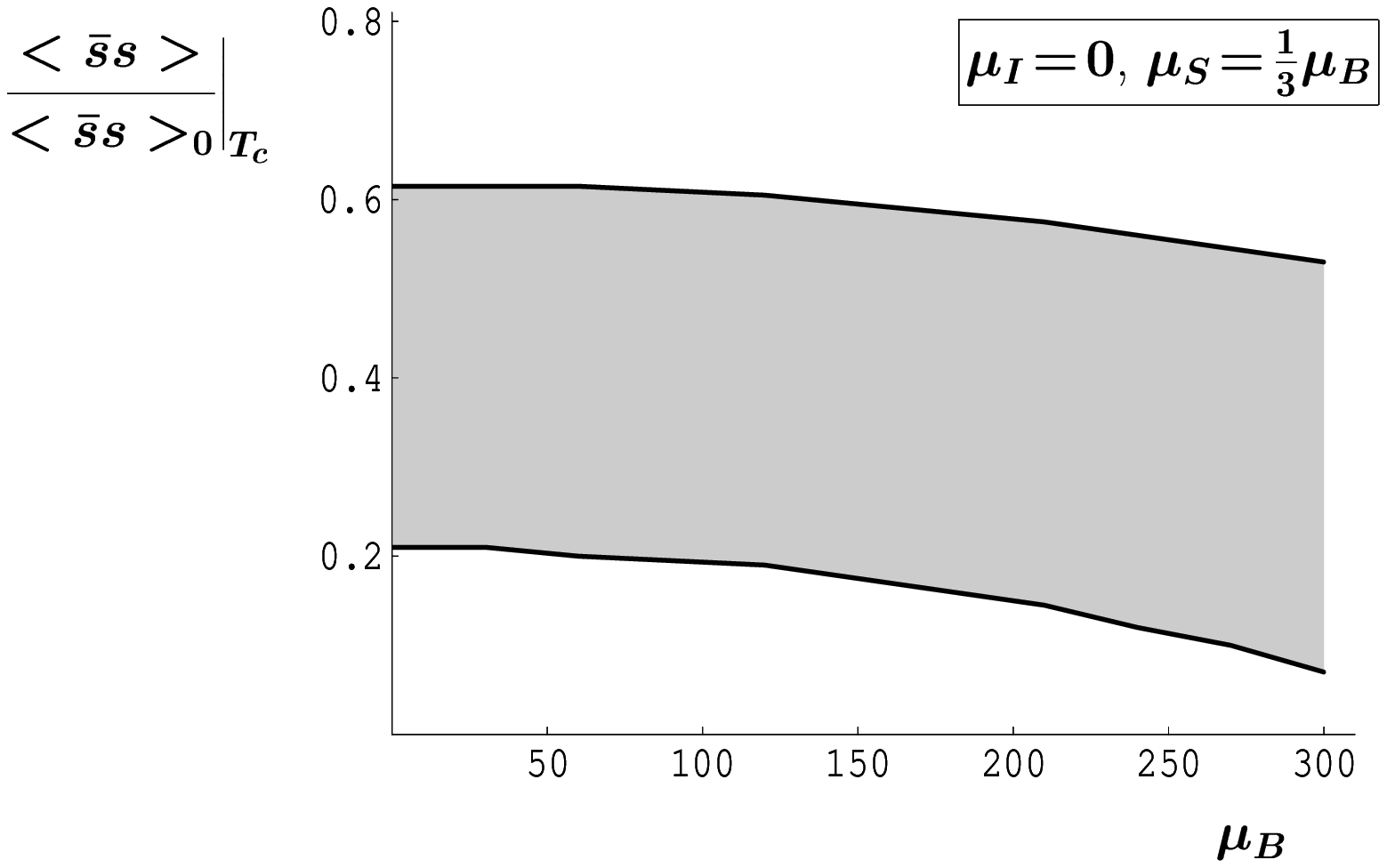}
\caption{\label{fig3}The value of $\frac{<\bar{s}s>}{<\bar{s}s>_0}$
  at the temperature where $<\bar{q}q>$ vanishes given as a function
  of baryon chemical potential $\mu_B$. On left panel, we assign $\mu_I$ and
  $\mu_S$ to zero. On right panel, we set $\mu_I=0$ and
  $\mu_S=\frac13\mu_B$. Latter values are relevant to the values measured
  in recent heavy-ion collisions.}  
\end{figure}

At zero chemical potential, we find that the strange quark-antiquark condensate
is still large at temperatures where the light quark-antiquark condensates
become small: $\frac{<\bar{s}s>}{<\bar{s}s>_0}=0.4\pm0.2$ where $<\bar{q}q>$
vanishes. As can be seen in Fig.~\ref{fig2}, an increase in $\mu_B$
tends to reduce the difference between $<\bar{q}q>$ and $<\bar{s}s>$.
Increasing $\mu_I$ apparently accentuates this difference. We also notice
that increasing $\mu_S$ does not sizably affect. The reason for this
behavior is that the light quark-antiquark condensates in the hadron
resonance gas model are much more sensitive to the pion physics than the
strange quark-antiquark one. On the other hand, we find that all
condensates are equally sensitive to $\mu_B$. 
The value of $\frac{<\bar{s}s>}{<\bar{s}s>_0}$ at the
critical temperature for the light quark flavors is shown in Fig.~\ref{fig3}
as a function of $\mu_B$. We study two cases: $\mu_I=0$ and $\mu_S=0$ and 
$\mu_I=0$ and $\mu_S=\mu_B/3$. The latter is very much compatible with
the corresponding values calculated for recent heavy-ion collision
experiments, as mentioned above. 

We conclude that the strange quark-antiquark condensate can be
relatively large at temperatures where the light quark-antiquark condensate
is very small. We obtained these results from a hadron resonance gas
model.  Our results agree with the recent MILC lattice
simulations at zero temperature and chemical potentials~\cite{MILCthermo}. The 
uncertainty of the value of the strange quark-antiquark condensate at this
critical temperature is however rather large. This is due to the
uncertainty in the values of the condensates at high temperatures which we
have calculated by the hadron resonance gas model. This uncertainty could
be reduced for the most part, if the factor $A$ in Eq.~(\ref{lattMass})
were more precisely known.

\section{\label{sec:4}Conclusion}

We have shown that in the hadron resonance gas model, the strange
quark-antiquark condensate can still be sizably large at temperatures where
the light quark condensates nearly vanish. In other words, the hadron
resonance gas model seems to indicate that the chiral symmetry could still
broken for the strange quark flavor at temperatures where chiral symmetry
is restored for the light quark flavors. This new situation seems to be
possible only in a range of a few tens of MeV around $T=180$~MeV. 
{\sf This result is in agreement with recent lattice calculation at zero
chemical potential. We have shown that this effect persists for small
chemical potentials and that it seems to slowly disappear at higher
chemical potentials.} We have also shown that this effect can be enhanced by
a nonzero isospin chemical potential.  This fact could be used in heavy-ion
collision experiments by choosing different isotopes, which is an
interesting option for other reasons as well \cite{muBI}. 

This effect does not contradict any fundamental principle. At
nonzero quark masses, lattice simulations show that in QCD 
the hadronic phase and the quark-gluon plasma phase are separated by a
crossover, not by a sharp phase transition. It is therefore possible for
the light quark-antiquark condensates to be small while the strange
quark-antiquark condensate is large in the crossover region.  However,
it is generally expected that this crossover stops and becomes a first
order phase transition with increasing $\mu_B$.  In this latter
case, $<\bar{q}q>$ and $<\bar{s}s>$ should drop and become small at the same
temperature. Furthermore, the discontinuity of $<\bar{q}q>$ and
$<\bar{s}s>$ at a first order phase transition are related by a
Clausius-Clapeyron relation: They should be similar~\cite{CCrel, largeN}.
The emergence of a first order phase transition at nonzero $\mu_B$ 
 is therefore entirely possible in the hadron resonance gas model, since
the difference between $<\bar{q}q>$ and $<\bar{s}s>$ is smaller at larger 
 $\mu_B$.  Notice however, that the difference between
$<\bar{q}q>$ and $<\bar{s}s>$ is larger for larger $\mu_I$. This would tend
to indicate that there is no critical endpoint in the QCD phase diagram at
nonzero temperature and isospin chemical potential.

Finally, we observe that if this effect is true,  our study opens the
possibility of the existence of a domain where chiral symmetry
is partially restored with a spontaneous chiral symmetry breaking given by
$SU(3)_L\times SU(3)_R \rightarrow SU(3)_V \times SU(2)_A$. It would
be interesting to study the physics of such a spontaneous symmetry
breaking of chiral symmetry and determine its consequences for heavy-ion
collision experiments.

\end{document}